# FERMILAB PXIE BEAM DIAGNOSTICS DEVELOPMENT AND TESTING AT THE HINS BEAM FACILITY*

V. E. Scarpine[#], B. M. Hanna, V. Lebedev, L. Prost, A. V. Shemyakin, J. Steimel, M. Wendt, Fermilab, Batavia, IL 60510 USA

## Abstract

Fermilab is planning the construction of a prototype front end of the Project X linac. The Project X Injector Experiment (PXIE) is expected to accelerate 1 mA CW H- beam up to 30 MeV. Some of the major goals of the project are to test a CW RFQ and H- source, a broadband bunch-by-bunch beam chopper and a low-energy superconducting linac. The successful characterization and operation of such an accelerator place stringent requirements on beamline diagnostics. These crucial beam measurements include bunch currents, beam orbit, beam phase, bunch length, transverse profile and emittance and beam halo and tails, as well as the extinction performance of the broadband chopper. This paper presents PXIE beam measurement requirements and instrumentation development plans. Presented are plans to test key instruments at the Fermilab High Intensity Neutrino Source (HINS) beam facility. Since HINS is already an operational accelerator, utilizing HINS for instrumentation testing will allow for quicker development of the required PXIE diagnostics.

## PXIE

Fermilab is planning a program of research and development aimed at integrated systems testing of critical components of the front end of the Project X linac [1]. The mission goals include (1) a platform for demonstrating the operation of Project X front end components at full design parameters, (2) the delivery of 1 mA average current with 80% bunch-by-bunch chopping from the RFQ and (3) efficient acceleration with minimal emittance dilution through at least 15 MeV beam energy. This integrated systems testing, known as the Project X Injector Experiment (PXIE), will be completed over the period FY12-16.

Figure 1 shows a block diagram of the proposed PXIE beamline. The main components of this beamline are

- CW H- ion source delivering 5 mA at 30 keV
- Low-Energy Beam Transport (LEBT) with pre-chopping
- CW Radio Frequency Quadrupole (RFQ) operating at 162.5 MHz and delivering 5 mA at 2.1 MeV
- Medium-Energy Beam Transport (MEBT) with integrated wide-band, arbitrary bunch pattern chopper and beam absorber
- Low-β superconducting Half-Wave Resonator (HWR) and Single-Spoke Resonator (SSR) cryomodules accelerating 1 mA beam to 30 MeV
- Beam dump up to 50 kW for extended periods.

Further information on PXIE can be found in other papers of this conference [2-3] and at the PXIE website [4].

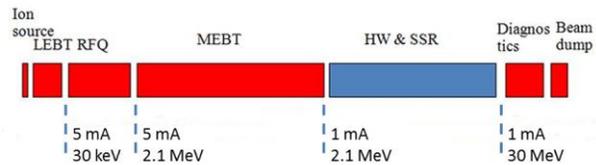

Figure 1: Proposed beamline for PXIE.

## BEAM DIAGNOSTICS FOR PXIE

The successful characterization and operation of PXIE place stringent requirements on beamline diagnostics. The following is a laundry list of diagnostics that is being investigated for use in PXIE.

- Beam current: DCCT, toroid, resistive wall current monitor
- Beam orbit: button-style BPMs, digital read-out electronics
- Beam energy and phase: BPMs, time-of-flight and spectrometer in dump beamline
- Beam emittance: Allison scanner, slit and wirescanner, laser-wire
- Longitudinal profile: mode-locked laser, Feschenko wire-style monitor, fast Faraday cup
- Chopper extinction monitor: resistive wall current monitor
- Transverse halo: vibrating wire, collimators
- Longitudinal tails: Feschenko wire monitor

A discussion of all of the devices is prohibitively long for this conference proceeding. We shall therefore focus on only a few of these diagnostic instruments.

### PXIE LEBT Beam Diagnostics

Figure 2 shows a block diagram of the proposed LEBT beamline for PXIE. The function of the LEBT is to transport and control the H- beam from the source to the entrance of the RFQ. The LEBT includes a low-bandwidth beam chopper that can provide up to 1 MHz of beam chopping. In anticipation of its operation as the Project X front-end, the PXIE LEBT can also accommodate a second H- ion source through the use of a switching dipole magnet.

The primary function of diagnostics in the LEBT is to characterize and tune the beam during operations. The quality of the beam out of the source will be monitored by periodically measuring the vertical and horizontal transverse emittance using a pair of Allison scanners. Because these scanners are upstream of the LEBT beam

___

*Work supported by the U. S. Department of Energy, contract No. DE-AC02-07CH11359
#scarpine@fnal.gov

chopper, they will need to be water-cooled in order to operate up to 300 KW of beam power.

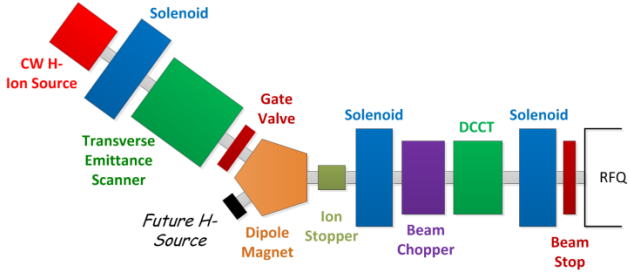

Figure 2: Proposed PXIE LEBT beamline.

One drawback of using Allison scanners for emittance measurements is that they will inhibit transport through the LEBT. Another possible drawback is that any beam hitting the face of one of the scanners will liberate a large number of electrons, possibly altering the beam emittance. For these reasons we are also investigating the possibility of utilizing a laser emittance monitor in the LEBT.

Beam current will be measured in the LEBT using a DC Current Transformer (DCCT). The DCCT is positioned after the LEBT chopper in order to allow it to measure both nominal DC beam as well as chopped beam.

## PXIE MEBT Beam Diagnostics

The primary functions of the PXIE MEBT are (1) to match optical functions between the RFQ and the superconducting HWR cryomodule and (2) to generate arbitrary bunch patterns at 162.5 MHz using an integrated wide-band chopper and beam absorbers, capable of disposing of 4 mA average beam current [5]. In addition, the MEBT will include beam diagnostics to measure the beam properties coming out of the RFQ and into the HWR cryomodule. Figure 3 shows a block diagram of the proposed PXIE MEBT indicating the location of matching, chopping and diagnostics sections.

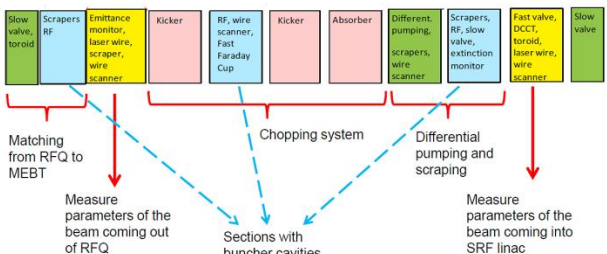

Figure 3: Proposed PXIE MEBT.

Two MEBT operating conditions put strict requirements on the choice of beam diagnostics. First, because of the nominal high beam power in the MEBT, beam-intercepting diagnostics can operate only with LEBT chopped beam. However, there is still danger of damage to the intercepting diagnostics contaminating the downstream superconducting cavities. To remedy this, we are investigating non-intercepting laser wire diagnostics for beam profiling. Second, to verify the operation of the arbitrary bunch wide-band chopper, bunch-by-bunch extinction must be measured to a level of $10^{-4}$.

## Laser Diagnostics for Beam Profiling

We are pursuing the use of lasers to obtain noninvasive measurements of both longitudinal and transverse profiles of $H^-$ beam via photo-detachment, $H^- + \gamma \rightarrow H^o + e^-$. The SNS accelerator has commissioned transverse beam profiling using a high-power Q-switched Nd:YAG laser [6]. In addition, they have also demonstrated longitudinal beam profiling using 2.5 ps laser pulses from a mode-locked Ti:Sapphire laser [7]. These systems used high peak laser power to saturate the photo-detachment process.

In general, these systems suffer from vibration and temperature drift issues due to the need to transport intense laser pulses from outside the beamline enclosure to the individual profile monitors[6]. To mitigate these issues, we are investigating the use of fiber optics to transport the lower power laser pulses. One technique would distribute low-power, amplitude modulated laser pulse trains via optical fibers. Since lower laser power will drive the photo-detachment process into saturation, synchronous detection will be used to measure the weaker detached electron signal. Figure 4 shows a block diagram of this setup using a 1 MHz amplitude modulation [8].

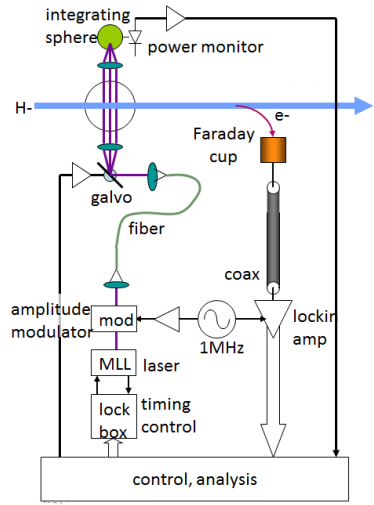

Figure 4: Fiber optic laser distribution profiling system with synchronous detection.

## MEBT Beam Chopper Extinction Diagnostics

In order to monitor the operation of the MEBT wide-band chopper, beam diagnostics will need to measure the bunch-by-bunch extinction level. PXIE plans on utilizing a wide-band Resistive Wall Current Monitor (RWCM) initially tested at EMMA [9]. The RWCM has a relatively flat frequency from 10 kHz up to ~ 4 GHz. This enables the RWCM to see individual bunches in the MEBT. Two RWCM signals will use fast integrators to measure the beam current before and after the chopper. The average bunch extinction level will be measured using a high-bandwidth oscilloscope similar to Fermilab Sampled

Bunch Displays [10]. Figure 5 shows a block diagram of the system in the PXIE MEBT.

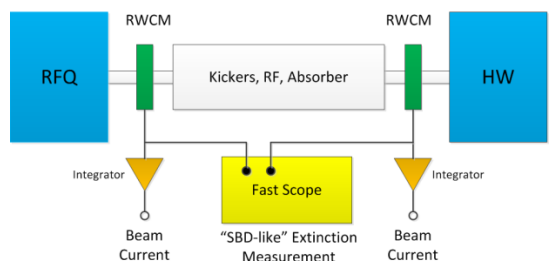

Figure 5: PXIE MEBT beam current and extinction measurements.

## BEAM DIAGNOSTICS DEVELOPMENT AT HINS

Fermilab initially developed a low-energy, high-intensity pulsed front-end test accelerator, called HINS, to pursue advanced linac technologies [11]. HINS has met its primary goal of demonstrating the use of high power RF vector modulators to control multiple RF cavities by a single high-power klystron for acceleration of a non-relativistic beam [12].

The present HINS configuration consists of a 50 keV proton source, a 2.5 MeV Radiofrequency Quadrupole (RFQ) followed by room-temperature spoke resonator acceleration section, and a beam diagnostics section. Nominal beam operation generates up to 8 mA of 3 GeV 200 μs proton pulses at a rate of 1 Hz. Figure 6 is a block diagram of the HINS beamline.

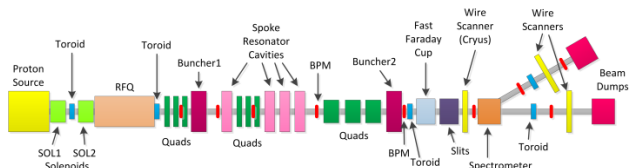

Figure 6: HINS beamline configuration.

The HINS is a unique linac injector R&D facility with access to high-intensity, low-energy beam for user projects. HINS provides an opportunity for the development of projects for PXIE and can be a facility for external collaborators. Potential project areas for PXIE beam diagnostics are
- BPM development
- Laser wire development for both transverse and longitudinal profiles and transverse emittance
- RWCM extinction measurements.

Figure 7 shows a proposed change to the HINS beamline for the development of PXIE laser wire diagnostics as well as RWCM extinction measurements. This configuration requires that the HINS ion source be changed to an H- source.

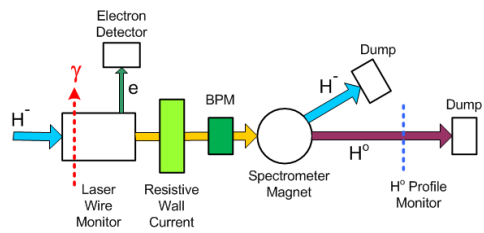

Figure 7: Proposed HINS station for PXIE diagnostic instrumentation development.